\documentclass[a4paper,11pt]{article}
\pdfoutput=1 

\usepackage{jinstpub} 

\usepackage{lineno}
\usepackage{caption}
\usepackage{subcaption}
\usepackage{wrapfig}
\usepackage[separate-uncertainty=true]{siunitx}
\usepackage[version=4]{mhchem}

\title{\boldmath A UV laser test bench for micro-pattern gaseous detectors}

\author[a,b,1]{A. Pellecchia,\note{Corresponding author.}}
\author[a]{A. Ranieri,}
\author[a]{P. Verwilligen}

\affiliation[a]{INFN sezione di Bari, IT}
\affiliation[b]{Università degli studi di Bari, IT}

\emailAdd{antonello.pellecchia@ba.infn.it}

\abstract{UV lasers for the calibration of gaseous detectors have seen newfound employment with current-generation micro-pattern gaseous detectors (MPGDs), especially those devices not suitable for tests with traditional sources (e.g. cosmic rays or X-ray beams) by design constraints. An apparatus made of a UV laser designed for the characterization of the fast timing MPGD (FTM) is here described, together with the measurements of gain curve and electron drift velocity performed on a prototype of Time Projection GEM (TPG) to validate the setup.}

\keywords{Gaseous detectors, Micropattern gaseous detectors, Time projection chambers, Lasers, Detector alignment and calibration methods}

\proceeding{15th Topical Seminar on Innovative Particle and Radiation Detectors\\
  14-17 October 2019\\
  Siena, Italy}

\begin{document}
\maketitle
\flushbottom


\section{Introduction}
\label{sec:intro}

The characterization of gaseous detectors for particle physics is commonly performed with high-energy radiation, such as cosmic rays, X-rays or particles in test beams; drawbacks of such techniques include sensitivity to magnetic fields and large energy deposit fluctuations (specifically for charged particles), low spatial accuracy and non-monochromatic source energy spectra (particularly for photon beams in the \SI{}{\kilo\eV} region) or the availability and large amount of work and organization required for testing detectors with test beams.

Laser rays can be used for the calibration of gaseous detectors, as they can provide an energy deposit along their tracks in typical gas mixtures suitable to simulate the passage of one to several MIPs \cite{hilke_lasers}. Characterization techniques with lasers have newfound utility in the current and next-generation \emph{micro-pattern gaseous detectors} (MPGDs), especially those made of fully resistive materials or particular design constraints. One such detector is the \emph{fast timing MPGD} (FTM), which aims to improve on the time resolution limit of current state-of-the-art MPGDs (set for instance to about 5-\SI{10}{\nano\s} for the triple-GEM chambers of the CMS experiment \cite{cms_gem_tdr}) by one order of magnitude by employing a multi-layer design \cite{ftm}.

\begin{figure}[h]
    \centering
    \begin{minipage}[c]{0.59\linewidth}
        \includegraphics[width=.8\textwidth]{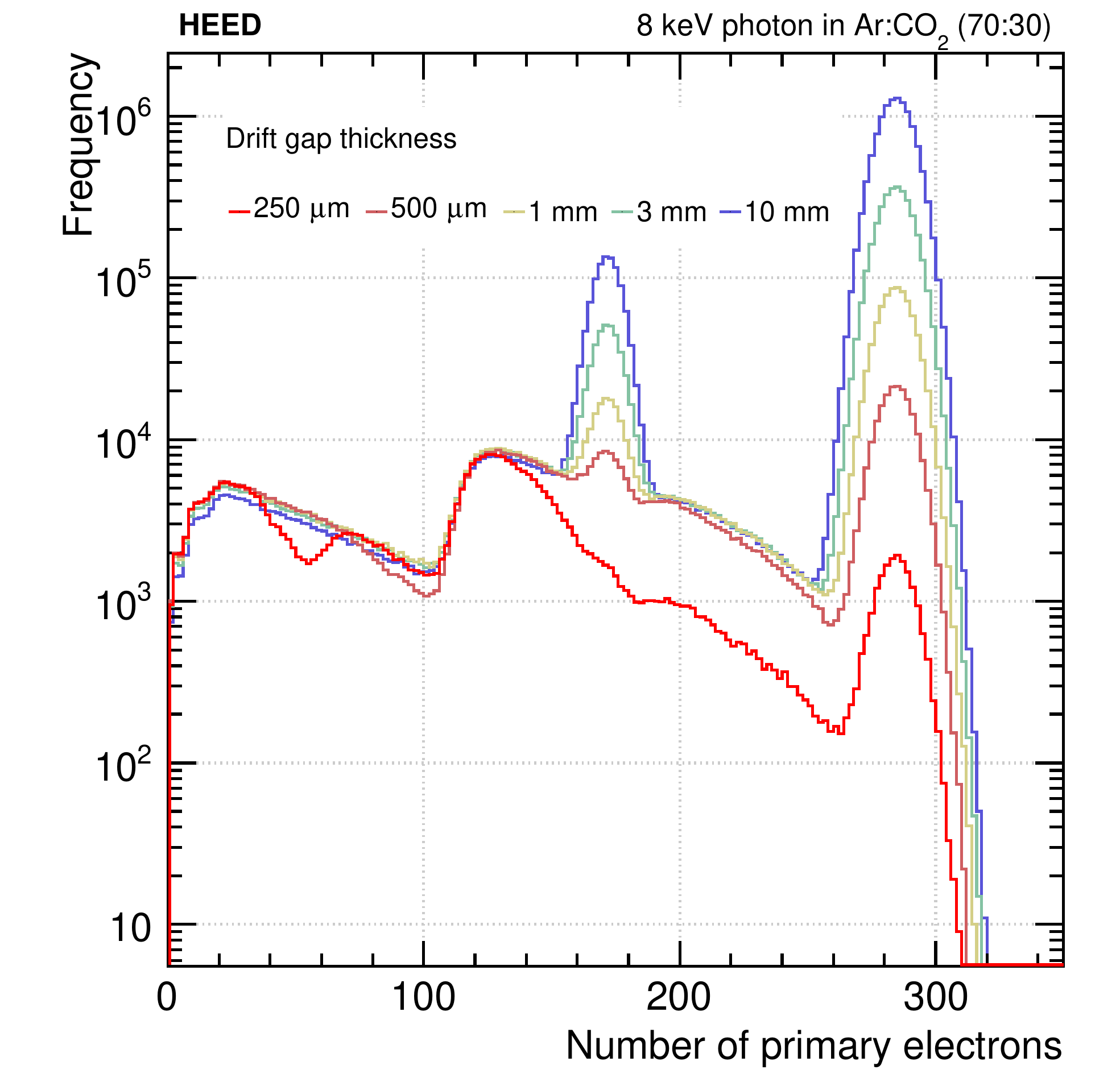}
    \end{minipage}
    \begin{minipage}[c]{0.4\linewidth}
        \caption{Distribution of the number of primary electrons released by an \SI{8}{\kilo\eV} photon in drift gaps of different thicknesses filled with Ar:\ce{CO2} mixture (70:30), simulated with \texttt{HEED}; the primary peak becomes non-distinguishable for very thin gaps. This effect makes the fluctuations on the energy deposit too large and renders X-rays ineffective for gain measurements.}
        \label{fig:heed_xray}
    \end{minipage}
\end{figure}

Current prototypes of the FTM, at the present moment under early development, feature exclusively resistive electrodes and thin drift gaps (as low as \SI{250}{\micro\m}), which prevent its gain measurement with the traditionally employed X-ray sources (either natural radioactive sources or X-ray tubes) because of two main reasons:
\begin{itemize}
    \item the low penetrating power of low-energy X-rays (less than \SI{10}{\kilo\eV}) in a multi-layer structure made of PCB material, combined with the lack of copper electrodes, which are typically employed to convert the non-monochromatic spectrum of higher-energy (more than \SI{10}{\kilo\eV}) sources to the characteristic monochromatic copper fluorescence spectrum (peaked at \SI{8.9}{\kilo\eV});
    \item in thin drift gaps, part of the energy released by photons in the \SI{}{\kilo\eV} energy region is lost without being converted to primary electrons in the gas, so the total number of primaries created by ionization by a single X-ray photon is subjected to large fluctuations (figure~\ref{fig:heed_xray}).
\end{itemize}

The laser facility described here has been designed for the characterization of the future FTM prototypes and tested with a triple-GEM detector for validation purposes.

\section{The laser setup}

\subsection{Laser-gas interaction}
\label{subsec:laser_gas_interaction}

Laser rays are known to produce ionization in common gas mixtures used to fill detector volumes by interaction with small-concentration impurity molecules, which are naturally present due to outgassing of the detector walls or the gas pipes \cite{bourotte}. Such molecules have low ionization potential (\SI{9}{\eV} or less) and can then be ionized in multi-photon processes by ultraviolet laser beams.

The rate of ionizations per unit volume due to $n-$photon absorption for an impurity species of molecule density $N$ (in \SI{}{\m^{-3}}) is
\begin{equation}
    \label{eq:ionization_density_1}
    \frac{R}{V} = N\sigma^{(n)}\phi^n,
\end{equation}
where $\phi$ is the flux of the laser beam (i.e. the number of photons crossing a unit area per unit time) and $\sigma^{(n)}$ is the \emph{$n-$photon cross-section equivalent}, measured in $\textrm{cm}^\textrm{2n}\textrm{s}^\textrm{n-1}$.

At low beam intensities, two-photon ionization dominates, with a ionization rate density
\begin{equation}
    \label{eq:ionization_density_2}
    \frac{R}{V} = \left(\frac{\lambda}{hc}\right)^2 N\sigma^{(2)}I^2,
\end{equation}
where $I=(hc/\lambda)\phi$ is the laser beam intensity and $A$ is the laser spot size at the ionization point. This equality can be obtained by solving the semi-classical rate equations for the population density of the impurity species \cite{blum}.

\subsection{Laser specifications}

The optical setup was made of a Nd:YAG pulsed laser of \SI{266}{\nano\m} wavelength (CryLaS 266-50), short enough to ionize molecules like benzene, toluene and cumene by two-photon interaction. The energy of a single laser pulse (controllable through an external attenuator) was \SI{51}{\micro\J} and the pulse duration was of 0.9-\SI{1.3}{\nano\s} FWHM, about one order of magnitude shorter than the de-excitation times of many hydrocarbon molecules. The waist radius was \SI{400+-100}{\micro\m}, so that the beam intensity at the laser head was \SI{101+-51}{\micro\J/\milli\m^2} -- known to be sufficient for a ionization yield equivalent to several minimum-ionizing particles \cite{alice_tpc}.

\begin{table}[ht]
    \centering
    \caption{Specifications of the CryLaS FQSS266-50 laser.}
    \label{table:laser}
    \begin{tabular}{|c|c|c|c|c|}
        \hline
        \bf{Pulse energy} & \bf{Waist radius} & \bf{Wavelength} & \bf{Pulse duration} & \bf{Spatial mode} \\
        \hline
        \SI{51}{\micro\J} & \SI{400}{\micro\m} & \SI{266}{\nano\m}/\SI{4.7}{\eV} & \SI{1}{\nano\s} FWHM & $\text{TEM}_{00}$ \\
        \hline
        can provide a & low angular & two-photon & lower than the & gaussian beam \\
        MIP-like energy & divergence & ionization of & TPG time & quality $<$1.5 \\
        deposit & & hydrocarbons & resolution & \\
        \hline
    \end{tabular}
\end{table}

\subsection{Preparation of the optical setup}

\begin{figure}[ht]
    \centering
    \begin{subfigure}[b]{\textwidth}
        \centering
        \includegraphics[width=.75\linewidth]{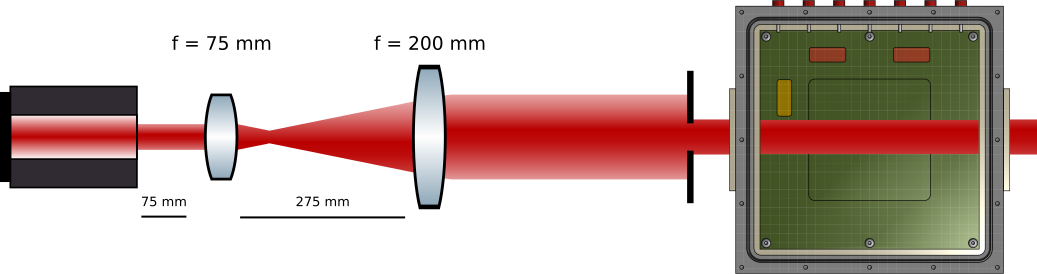}
        \caption{Collimated, low-intensity optical setup.}
        \label{fig:optical_setup_collimated}
    \end{subfigure}
    \par~\par
    \begin{subfigure}[b]{\textwidth}
        \centering
        \includegraphics[width=.75\linewidth]{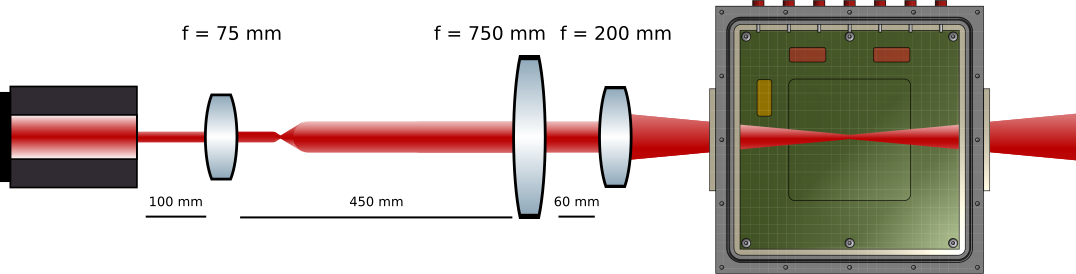}
        \caption{Focused, high-intensity setup.}
        \label{fig:optical_setup_focused}
    \end{subfigure}
    \caption{Optical setups designed for the laser box.}
    \label{fig:optical_setups}
\end{figure}

Two different optical setups (figure \ref{fig:optical_setups}) were designed for the laser bench, corresponding to a collimated beam at low intensity and a focused, high intensity beam.

In the first setup (figure \ref{fig:optical_setup_collimated}), the beam was collimated by a beam expander made of two lenses (focal lengths \SI{75}{\milli\m} and \SI{200}{\milli\m}) to an output waist radius of \SI{1500}{\micro\m}, with an angular divergence of \SI{0.06}{\milli\radian}. This had the effect of decreasing the overall intensity of the beam to \SI{34}{\micro\J\per\milli\m^2}; additionally, a pinhole was used to cut the outer part of the wavefront to lower the total pulse energy.

In the second setup (figure \ref{fig:optical_setup_focused}), the laser beam was focused by three lenses (75, 750 and \SI{200}{\milli\m} focal lengths respectively) down to a waist radius of \SI{23.4}{\micro\m}, in order to obtain a narrow point of primary ionization in the detector gas. At full pulse energy, the beam intensity in the waist in this configuration was \SI{3e4}{\micro\J\per\milli\m^2}.

\section{Characterization of the time projection GEM in the laser box}

\subsection{The time projection GEM prototype}

\begin{figure}[h]
    \centering
    \includegraphics[width=.45\linewidth]{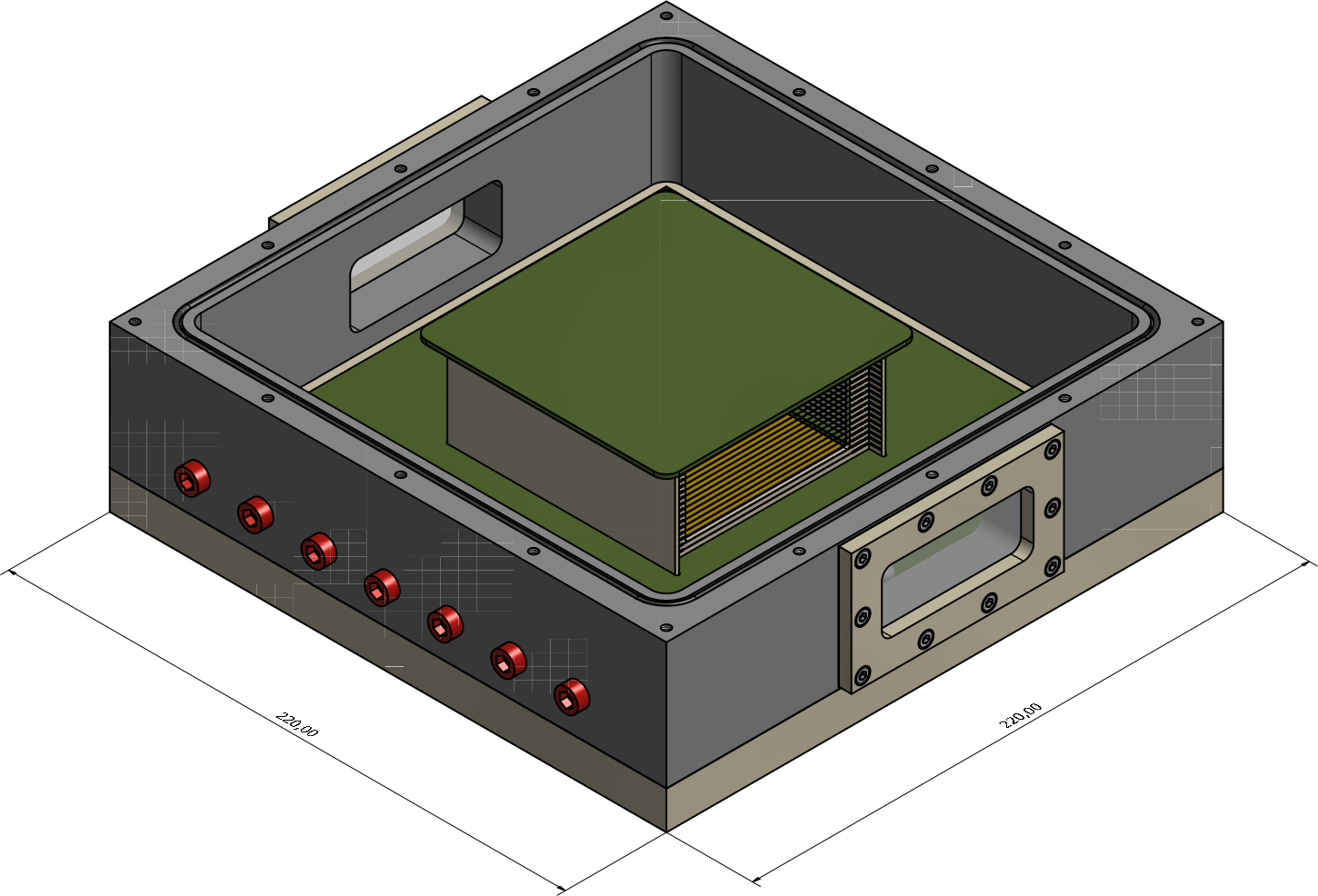}
    \includegraphics[width=.45\linewidth]{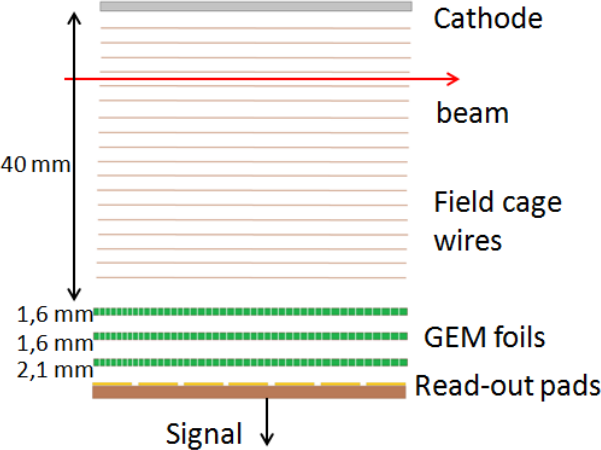}
    \caption{Design and gaps of the time projection GEM prototype.}
    \label{fig:tpg}
\end{figure}

\begin{figure}[h]
    \centering
    \begin{minipage}[c]{.49\linewidth}
        \centering
        \includegraphics[width=.6\textwidth]{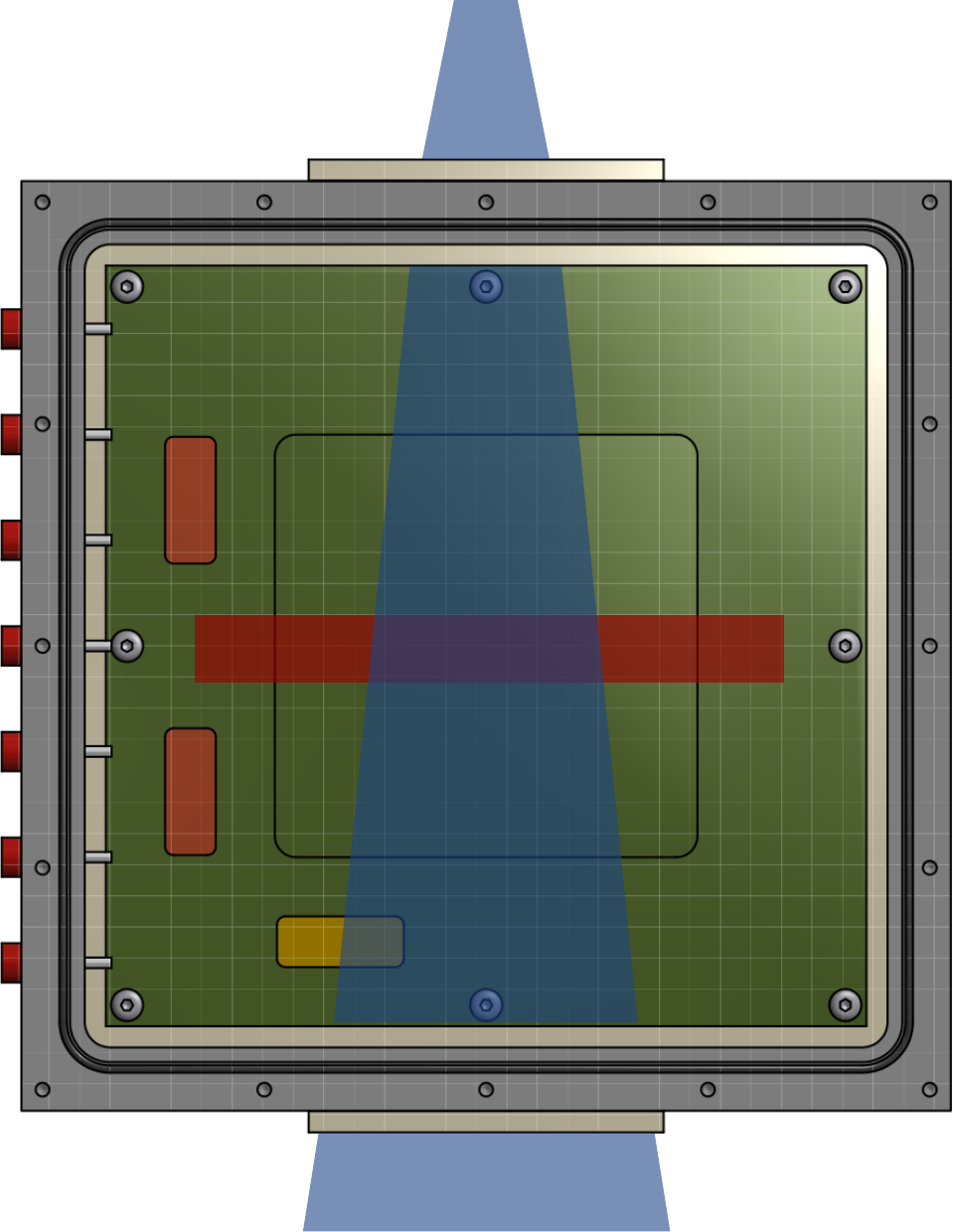}
    \end{minipage}
    \begin{minipage}[t]{.49\linewidth}
        \caption{Top projection of the TPG prototype. The blue shade represents the beam passing through the windows, while the red rectangle shows the instrumented area (corresponding to the segmented region of the anode).}
        \label{fig:tpg_2d}
    \end{minipage}
\end{figure}

The benchmark detector used to validate the laser box is a prototype of time projection chamber with GEM readout (TPG), i.e. a triple-GEM chamber with a \SI{40}{\milli\m} drift gap (figure \ref{fig:tpg}), originally designed for beam monitoring in hadron therapy \cite{altieri_tpg}. The beam enters the gas through one of the two opposite quartz windows, while the readout signals can be read through a dedicated electronics \cite{ciciriello} from the segmented region of the anode, made of two rows of $6\times\SI{2}{\milli\m^2}$ pads (figure \ref{fig:tpg_2d}).

Because of its small instrumented area compared to the total gas volume, the TPG is not suited for a gain calibration with an X-ray source (the \SI{8.9}{\kilo\eV} photon beam created by copper fluorescence is strongly attenuated while traveling in the gas before reaching the center of the detector).

\subsection{Primary ionization and gain measurement}
\label{subsec:gain_tpg}

To measure the gain of the TPG, the laser was operated at a \SI{100}{\Hz} repetition rate in the low-intensity, collimated setup (figure \ref{fig:optical_setup_collimated}); the detector was filled with Ar:\ce{CO2} mixture (70:30) and the electric fields in the gaps were fixed at the values in table \ref{table:fields_laser}, while the voltages on the GEM foils varied between 300 and \SI{400}{\V}.

\begin{table}[h]
    \centering
    \caption{Electric fields at which the TPG was operated in the laser box measurements.}
    \label{table:fields_laser}
    \begin{tabular}{ |c|c|c|c|c| } 
        \hline
        \bf Gap & Drift & Transfer 1 & Transfer 2 & Induction \\
        \hline
        \bf Field (\SI{}{\kilo\V/\centi\m}) & 0.4 & 3 & 3 & 3.3 \\
        \hline
    \end{tabular}
\end{table}

\begin{figure}
    \centering
    \includegraphics[width=.45\textwidth]{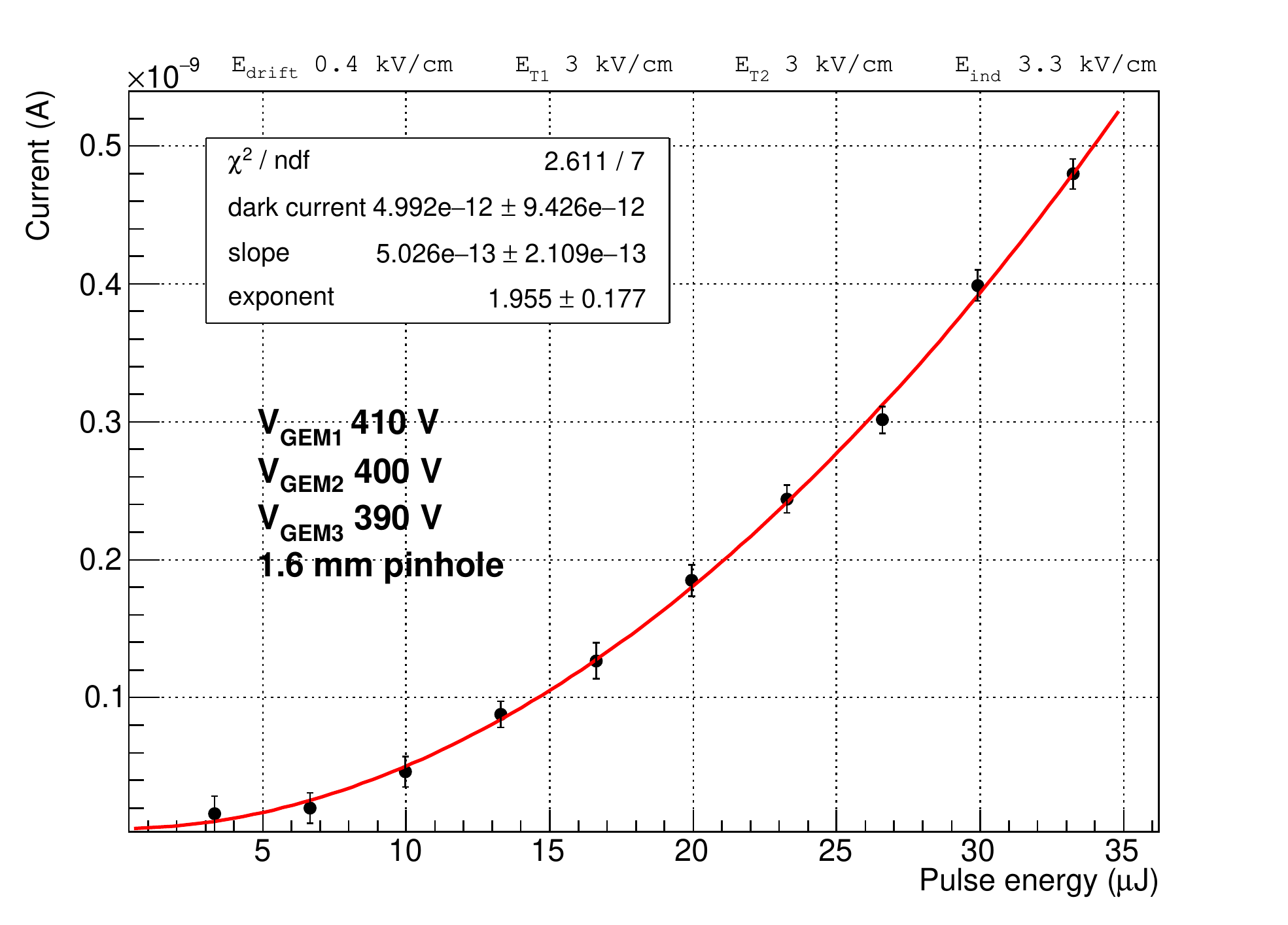}
    \centering
    \includegraphics[width=.45\textwidth]{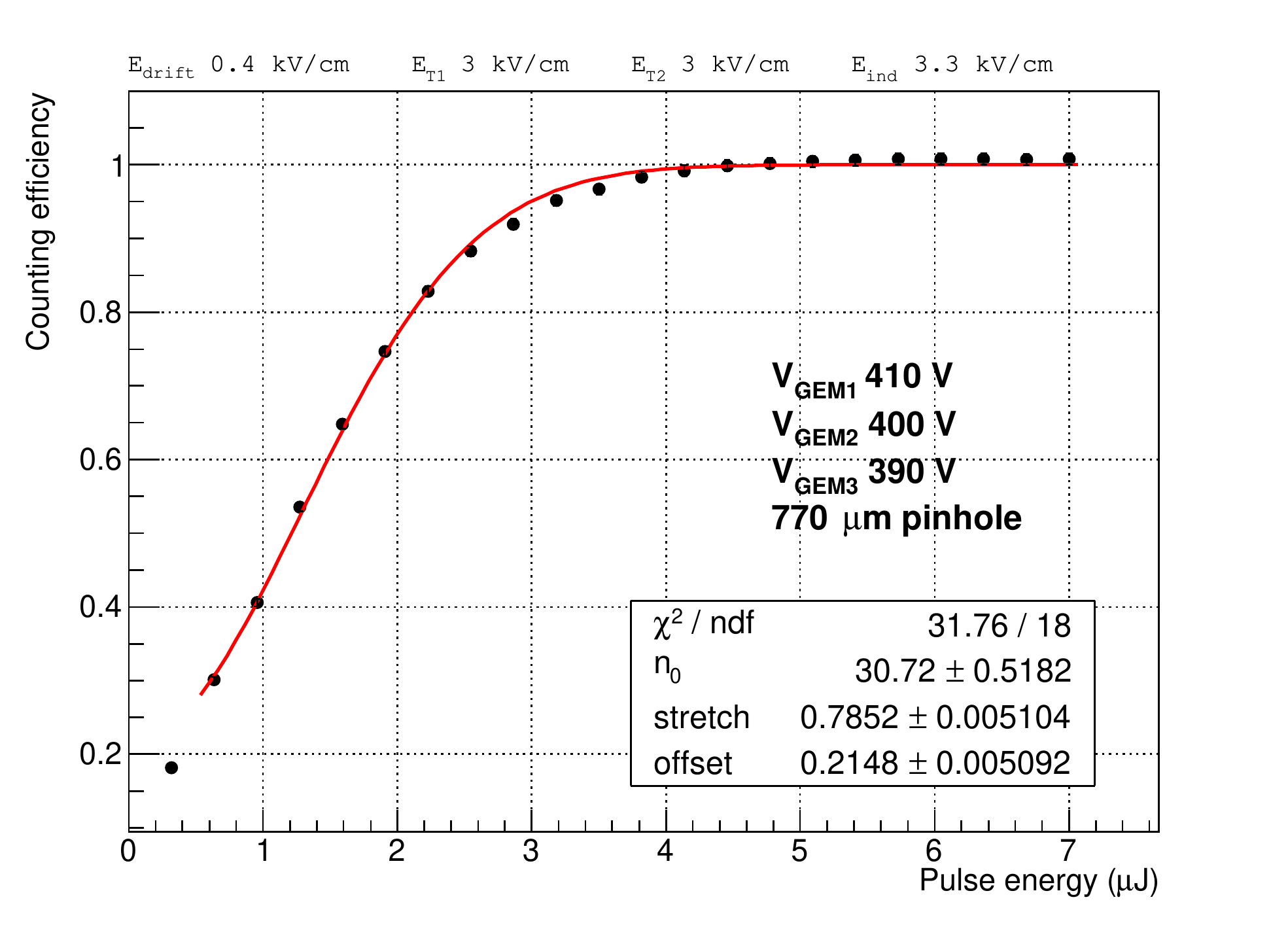}
    \caption{On the left, anode current measured from the TPG at different laser pulse energies. On the right, counting efficiency of the TPG, measured as in equation \ref{eq:counting_efficiency_def} and fitted with the function in equation \ref{eq:counting_efficiency_fit}.}
    \label{fig:tpg_current_and_counting_efficiency}
  \end{figure}

A direct measurement of the anode current of the TPG as a function of the laser pulse energy shows a quadratic behaviour (figure \ref{fig:tpg_current_and_counting_efficiency} shows the current measurement performed with a \SI{1600}{\micro\m} diameter pinhole). Since the anode current is proportional to the rate of primary ionization, this is a confirmation that two-photon absorption is the dominant ionization process, according to equation \ref{eq:ionization_density_2}.

Already established techniques for obtaining the gain curve of a detector with a laser setup involve observing the single-electron response \cite{zerguerras} or directly measuring the primary ionization current. In the technique followed here the effective gain of the TPG was determined as the ratio
\begin{equation}
    G_{\text{eff}} = \frac{\text{anode current}}{\text{primary current}} = \frac{\text{anode current}}{n_p\times q_e\times \text{laser rate}},
    \label{eq:effective_gain}
\end{equation}
where $q_e$ is the electron charge and $n_p$ is the number of primary electrons released in the gas by a single laser pulse at some fixed beam intensity.

The anode current was directly measured with a picoammeter, while the number of primaries $n_p$ was obtained from a scan of the signal \emph{counting efficiency} at the readout at different laser pulse energies. The counting efficiency is defined as the ratio
\begin{equation}
    \epsilon = \frac{\text{anode signal rate}}{\text{laser pulse rate}}.
    \label{eq:counting_efficiency_def}
\end{equation}

The resulting curve (figure \ref{fig:tpg_current_and_counting_efficiency}, measurement performed with a \SI{770}{\micro\m} diameter pinhole) is fitted by the function
\begin{equation}
    \epsilon =
    1 - 
    \sum_{n=0}^{n_{th}}\, \frac{\exp\,[n_0(E/E_0)^2]}{n!}\,
    n_0^n(E/E_0)^{2n},
    \label{eq:counting_efficiency_fit}
\end{equation}
where the fit parameters are the number $n_0$ of primary electrons created by a single laser pulse at some reference energy $E_0$ and the number $n_{\text{th}}$ of primary electrons corresponding to the voltage threshold set for the discriminator used in the signal count. The derivation of equation \ref{eq:counting_efficiency_fit} (obtained assuming Poisson fluctuations on the ionization rate) is recounted in the appendix.

From the fit, one obtains that the number of primary electrons created by a laser pulse of \SI{10}{\micro\J} is \SI{30.7+-0.5}{} over the \SI{12}{\milli\m} length of the instrumented area of the detector. This value is then used to estimate the primary ionization current, from which the effective gain is retrieved according to equation \ref{eq:effective_gain}.

\begin{figure}
    \centering
    \includegraphics[width=.8\textwidth]{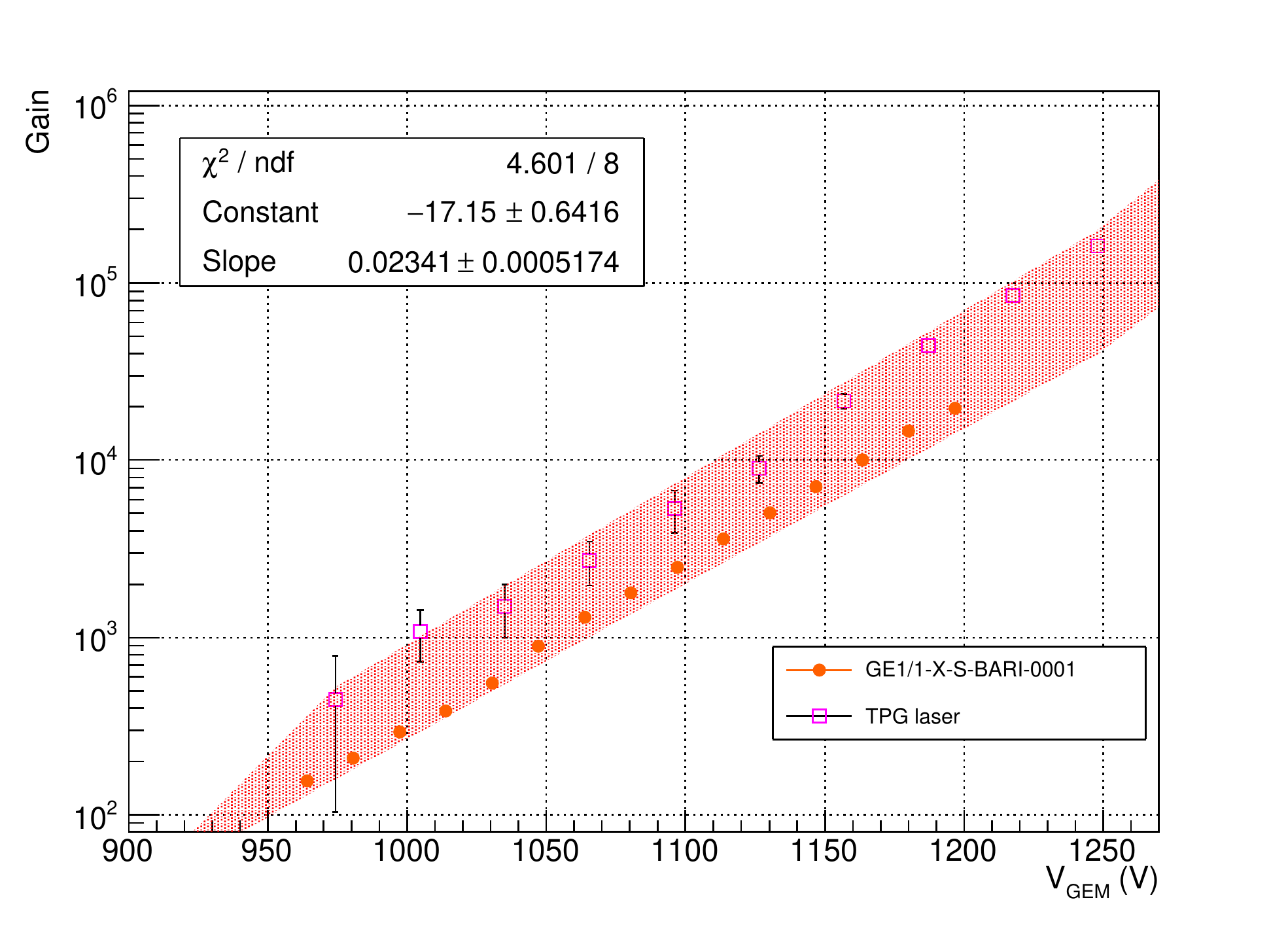}
    \caption{Measured effective gain of the TPG prototype as a function of the sum of the voltage differences on the three GEM foils (corrected for temperature and pressure); the red band shows the interval of gain curves of various triple-GEM chambers of the CMS experiment used as reference.}
    \label{fig:tpg_gain}
\end{figure}

The resulting gain curve is shown in figure \ref{fig:tpg_gain}; the comparison shows compatibility with the gain values of different triple-GEM chambers produced at CERN and the INFN section of Bari for the GE1/1 station of the CMS experiment.

\subsection{Timing measurements on the TPG}

The focused beam setup shown in figure \ref{fig:optical_setup_focused} was designed for measurements of time resolution and electron drift velocity in the TPG; the beam waist can be considered narrow enough for the ionization position in the detector gas to be practically point-like. The signal arrival time at the detector readout is dominated by the drift times of the electrons in the drift gap; therefore, the average electron velocity was determined by measuring the signal arrival times at different points of primary ionization inside the gas.

\begin{figure}
    \centering
    \includegraphics[width=.6\textwidth]{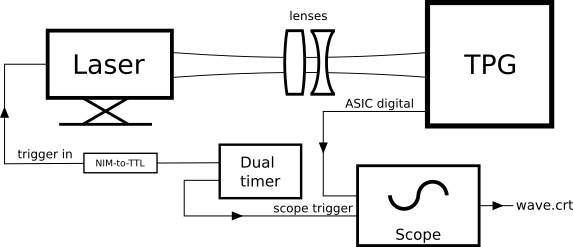}
    \caption{Acquisition setup for the signal arrival time measurement.}
    \label{fig:setup_time_resolution}
\end{figure}

Figure \ref{fig:setup_time_resolution} shows the electronic setup for the arrival time measurements; a dual timer acts as external trigger for both the laser and the fast scope that acquires and stores the detector signals for offline analysis. 

\begin{figure}
    \centering
    \includegraphics[width=.8\textwidth]{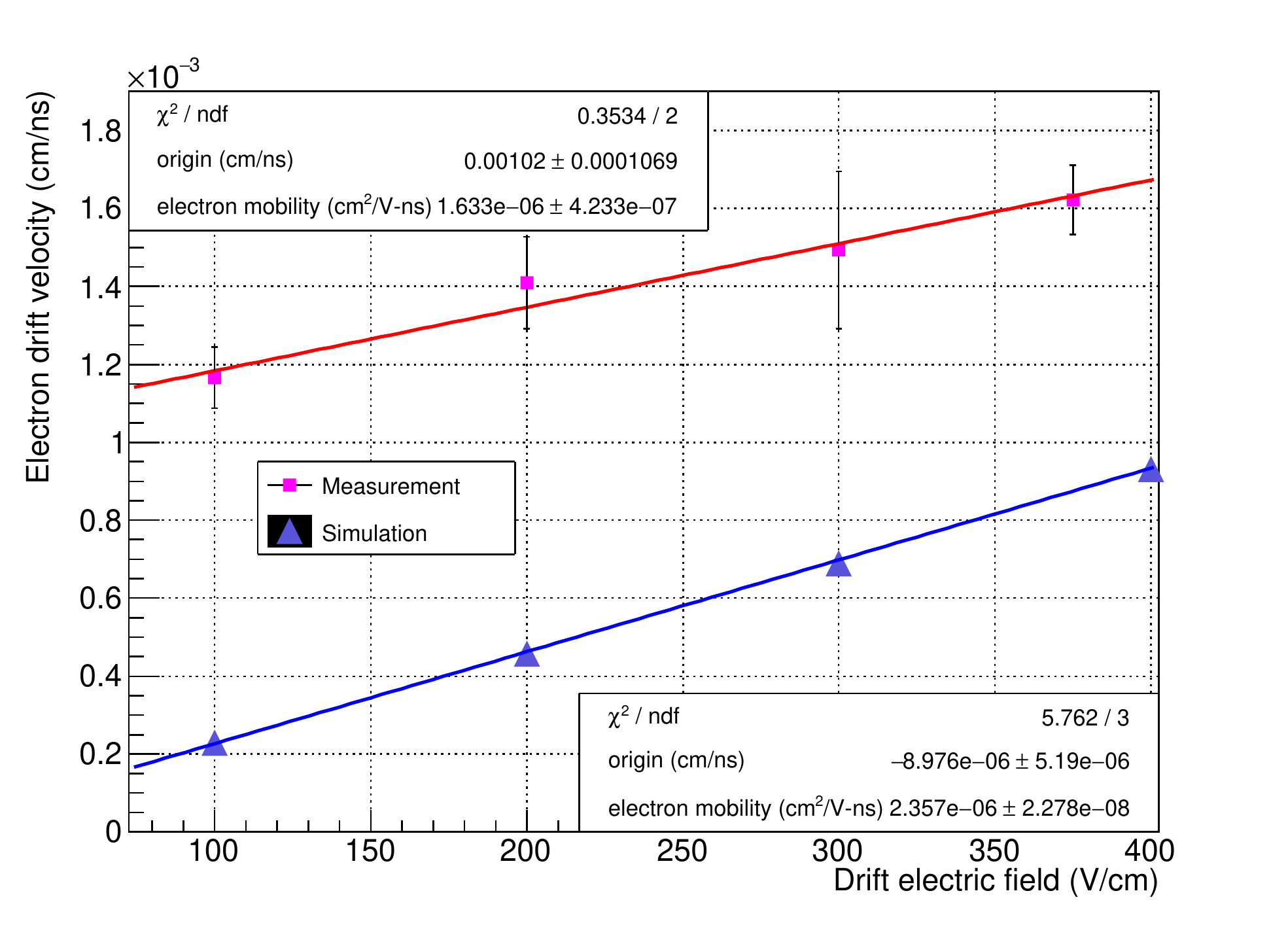}
    \caption{Electron drift velocity as a function of the electric field in the drift gap, compared with a \texttt{Magboltz} simulation.}
    \label{fig:electron_mobility}
\end{figure}

Figure \ref{fig:electron_mobility} shows the electron drift velocity measured at different electric fields. The estimated electron mobility is \SI{1.63+-0.42e-6}{\cm^2/\V\cdot\nano\second}, comparable with the value expected from a Monte Carlo \cite{magboltz} simulation (\SI{2.36e-6}{\cm^2/\V\cdot\nano\second}). The measured velocity curve presents an offset in the velocity curve for zero electric field with respect to the simulation, which might be due to the non-uniform electric field near the GEM holes; however, this need further investigation to clarify its origins without doubts.

\section{Conclusions}

The laser test bench has proved to be a viable setup for the calibration of MPGDs with particular design constraints. Its immediate future application is the full characterization of the FTM prototypes, for which the measurement procedures here described for the TPG will be employed. In particular, the gain calibration technique recounted in section \ref{subsec:gain_tpg} will be confirmed by cross-checking with a direct measurement of the primary ionization current; such measurement is indeed more feasible with a single-GEM detector than with the TPG, because of the generally lower collection efficiency of triple-GEM chambers.

\appendix
\section{Derivation of the interpolating function for the counting efficiency}

The counting efficiency as defined in equation \ref{eq:counting_efficiency_def} was measured by counting the readout signals of the TPG with a scaler preceded by a discriminator; the counting efficiency is then the probability for the readout signal to overcome the charge threshold set in the discriminator; for a fixed detector gain, such discriminator threshold corresponds to some threshold $n_\textrm{th}$ for the number of primary electrons created in the drift gap. Therefore,
\begin{equation}
    \epsilon = \frac{\text{anode signal rate}}{\text{laser pulse rate}} = \text{P}\{n_\textrm{electrons}\ge n_\textrm{th}\},
\end{equation}
where $n_\textrm{electrons}$ is the number of primary electrons created by a single pulse. Since the number of primaries is Poisson distributed,
\begin{equation}
    \epsilon =
    \sum_{n=n_{th}}^\infty \frac{\langle n\rangle^n}{n!}\text{e}^{-\langle n\rangle} =
    1 - \sum_{n=0}^{n_{th}} \frac{\langle n\rangle^n}{n!}\text{e}^{-\langle n\rangle}.
\end{equation}
As mentioned in section \ref{subsec:laser_gas_interaction}, the average number of primaries per pulse is proportional to the square of the beam energy; if $n_0$ is the average number of electrons created by a beam of fixed energy $E_0$, 
\begin{equation}
    \langle n\rangle = kE^2 = \frac{n_0}{E_0^2}\,E^2 = n_0\left(\frac{E}{E_0}\right)^2.
\end{equation}
The counting efficiency is then
\begin{equation}
    \epsilon =
    1 - 
    \sum_{n=0}^{n_{th}}\, \frac{\exp\,[n_0(E/E_0)^2]}{n!}\,
    n_0^n(E/E_0)^{2n}.
\end{equation}
In principle, the parameter $n_{th}$ is not known if there is no information on the detector gain. However, at first approximation it can be set to 0 and the fitting function becomes
\begin{equation}
    \epsilon = 1 -
    \text{e}^{n_0(E/E_0)^2}.
\end{equation}
The parameter $n_0$ can then be estimated by interpolation once an arbitrary reference energy $E_0$ is chosen, as done in figure \ref{fig:tpg_current_and_counting_efficiency}. The stretch and offset fitting parameters are introduced to account for the dark current.

\acknowledgments

The authors would like to thank INFN for the funding of this research
through the starting grant for young researchers bando 19593/2017.


\end{document}